\documentclass[aps,prl,twocolumn,groupedaddress,floatfix,showpacs,letterpaper]{revtex4}
\pdfoutput=1
\usepackage{times}
\usepackage[T1]{fontenc}
\usepackage{amssymb}
\usepackage{amsmath}
\usepackage{graphicx}
\usepackage{hyperref}

\graphicspath{{../graphs/}}
\usepackage{placeins}
\hypersetup{
    pdfauthor={Nathan Clisby},
    pdftitle={Accurate estimate of nu via a fast implementation of
      the pivot algorithm},
    breaklinks = true,
    colorlinks = true,
    linkcolor = blue,
    anchorcolor = blue,
    citecolor = blue,
    filecolor = blue,
    urlcolor = blue
    }
\providecommand{\e}[1]{\ensuremath{\times 10^{#1}}}

\newcommand{\Z}{{\mathbb Z}}

\begin{document}

\title{Accurate estimate of the critical exponent $\nu$ for
  self-avoiding walks via a fast implementation of the pivot
  algorithm\footnote{Journal reference: {\em Phys. Rev. Lett.},
    104:055702, 2010. doi: \href{http://dx.doi.org/10.1103/PhysRevLett.104.055702}{10.1103/PhysRevLett.104.055702}.}} 

\author{Nathan Clisby}
\email[]{n.clisby@ms.unimelb.edu.au}
\affiliation{ARC Centre of Excellence for Mathematics and Statistics of Complex Systems,
Department of Mathematics and Statistics, \\
The University of Melbourne, Victoria 3010, Australia}

\date{February 1, 2010}

\begin{abstract}
We introduce a fast implementation of the pivot algorithm for
self-avoiding walks, which we use to obtain large samples of walks on
the cubic lattice of up to $33\e{6}$ steps. Consequently the critical
exponent $\nu$ for three-dimensional self-avoiding walks is determined to
great accuracy; the final estimate is $\nu=0.587\; 597(7)$. The method can
be adapted to other models of polymers with short-range interactions,
on the lattice or in the continuum.
\end{abstract}

\pacs{64.60.De, 68.35.Rh, 64.70.km, 05.10.-a}

\maketitle

The self-avoiding walk (SAW) model is an important model in
statistical physics~\cite{Madras1993Books}. It models the excluded-volume
effect observed in real polymers, exactly capturing universal features
such as critical exponents. It is also the $n \rightarrow 0$ limit of
the $n$-vector model, which includes the Ising model ($n = 1$) as
another instance, thus serving as an important model in the study of
critical phenomena. 
Exact results are known for self-avoiding walks in two
dimensions~\cite{Nienhuis1982SAW,Lawler2004SAW} and
for $d \geq 4$ (mean-field behavior has been proved for $d \geq
5$~\cite{Hara1992SAW}), but not for the most physically interesting case of
$d=3$.

We have efficiently implemented the pivot algorithm
via a data structure we call the SAW-tree, which
allows rapid Monte Carlo simulation of SAWs of millions of
steps. We discuss this implementation in general terms here, and then
use 
this implementation to accurately calculate the critical exponent
$\nu$ for $\Z^3$. More details about the implementation can be found
in a companion article~\cite{Clisby2009bSAW}.
This new algorithm can also be adapted to other models of
polymers with short-range interactions, on the lattice and in the
continuum, and hence promises to be widely useful.

An $N$-step SAW on $\Z^d$ is a mapping
$\omega : \{0,1,\ldots, N\} \to \Z^d$ with $|\omega(i+1)-\omega(i)|=1$
for each $i$
($|x|$ denotes the Euclidean norm of $x$),
and with $\omega(i) \neq \omega(j)$ for all $i \neq j$.
We generate three-dimensional SAWs via the pivot algorithm, and
calculate various observables which characterize the size of the
SAWs: the squared end-to-end distance $R_{\mathrm{e}}^2$,
the squared radius of gyration $R_{\mathrm{g}}^2$, and the mean-square
distance of a
monomer from its endpoints $R_{\mathrm{m}}^2$, where
\begin{align*}
R_{\mathrm{e}}^2 &= |\omega(N)-\omega(0)|^2, \\
R_{\mathrm{g}}^2 &=
\frac{1}{2(N+1)^2}\sum_{i,j=0}^{N}|\omega(i)-\omega(j)|^2, \\
R_{\mathrm{m}}^2 &= \frac{1}{2(N+1)}\sum_{i=0}^{N}\left[|\omega(i)-\omega(0)|^2 +
  |\omega(i)-\omega(N)|^2 \right].
\end{align*}
We seek to calculate the mean values of these observables over all
SAWs of $N$ steps, where each SAW is given equal weight. Their
asymptotic forms are expected to be described by
\begin{align}
\langle R_x^2 \rangle_N &= D_x N^{2\nu}\left[1 + \frac{a_1}{N} + \frac{a_2}{N^2}
  + \cdots \right.\nonumber \\
&\left. + \frac{b_x}{N^{\Delta_1}} + \frac{b_1}{N^{2\Delta_1}} +
  \cdots + \frac{c_0}{N^{\Delta_2}} + \cdots \right] + \mathrm{af},
\label{eq:asymptotic}
\end{align}
with $0 < \Delta_1 < \Delta_2 < \cdots$, and where additional terms of the
form $c/N^{k_0+k_1\Delta_1+k_2\Delta_2+k_3\Delta_3+\cdots}$
($k_0,k_1,k_2,k_3,\cdots \geq 0$)
are not shown.
In addition, $\mathrm{af}$ indicates terms arising from the
anti-ferromagnetic singularity, which occurs in models on loose-packed
lattices such as $\Z^d$; these terms are negligible compared with
terms included in fits.
The exponents $\nu$, $\Delta_1$, and $\Delta_2$ are
universal, i.e. they are dependent only on the
dimensionality of the lattice and the universality class of the model,
while the amplitudes $D_x$ are observable dependent. However,
amplitude ratios, such as 
$D_{\mathrm{g}}/D_{\mathrm{e}}$ and $b_{\mathrm{g}}/b_{\mathrm{e}}$,
are universal quantities.

The pivot algorithm is a powerful approach to the
study of self-avoiding walks, invented by
Lal~\cite{Lal1969SAW} and later elucidated and popularized by Madras and
Sokal~\cite{Madras1988SAW}.
From an initial SAW of length $N$, such as a straight rod, new
$N$-step walks
are successively
generated by choosing
a site of the walk at random, and attempting to apply a lattice
symmetry operation,
or pivot, to one of the parts of the walk; if the
resulting walk is self-avoiding the move is
accepted, otherwise the move is rejected and the original walk is
retained.
The group of lattice symmetries for $\Z^3$ has 48 elements, and we
use all of them except the identity as potential pivot operations;
other choices are possible.
Thus a Markov chain is formed in the
ensemble of SAWs of fixed length;
this chain satisfies detailed balance and is ergodic,
ensuring that SAWs are sampled uniformly at random. Furthermore, as
demonstrated by Madras and Sokal~\cite{Madras1988SAW} through strong
heuristic arguments
and numerical experiments, the Markov chain
has a short integrated autocorrelation time for global observables,
thus making the pivot algorithm extremely efficient in comparison to
Markov chains utilizing local moves. See \cite{Madras1988SAW,Li1995SAW} for
detailed discussion.

The implementation of Madras and Sokal utilized a hash table to record
the location of each site of the walk. They showed that the pivot
algorithm has integrated autocorrelation time $O(N^p)$, with $p$
dimension-dependent but close to zero ($p \lesssim 0.2$), and argued
convincingly that the CPU time per successful 
pivot is $O(N)$ for their implementation.

Madras and Sokal argued that $O(N)$ is best possible because it takes
time of order $N$ to merely write down an $N$-step SAW.
However, Kennedy~\cite{Kennedy2002aSAW} recognized that it is \emph{not}
necessary to 
write down the SAW for each successful pivot, and from clever use of
geometric constraints developed an algorithm that broke the $O(N)$
barrier. The CPU time for this implementation grows as a
dimension-dependent fractional power of $N$ (see
Table~\ref{tab:performance}). 

\section{Method}

We have extended this idea to obtain a radical further improvement:
for $\Z^2$ and $\Z^3$ the mean CPU time per attempted pivot, which we
denote $T(N)$, is now only $O(\log N)$ for the range of $N$ studied,
and we have a theoretical argument that the large $N$ behavior is
$O(1)$.
The key observation is that although there are typically $O(N)$ nearest
neighbor contacts for a SAW of length $N$, the
number of contacts between two halves of a SAW is typically $O(1)$, as shown
via renormalization group~\cite{Muller1998SAW} and
Monte Carlo~\cite{Baiesi2001SAW} methods. When we attempt to pivot part
of a SAW, it is guaranteed that each of the two sub-walks remain
self-avoiding, and hence we only need to determine if
the sub-walks intersect. If the resulting walk is self-avoiding,
then we expect, on average, that there will be a constant number
of contacts between the
two sub-walks.

We will now briefly discuss the relevant data structure and
algorithms; full details can be found in \cite{Clisby2009bSAW}.
We implement a binary tree data structure (see
e.g. \cite{Sedgewick1998Books}) which we call a
\emph{SAW-tree}.
The root node of the SAW-tree contains information
about the whole walk, including $R_{\mathrm{e}}^2$,
$R_{\mathrm{g}}^2$, $R_{\mathrm{m}}^2$, and its 
{\em minimum bounding box}, which
is the smallest rectangular prism with faces of the form $x_i = c$
which completely contains the walk. The two children of the root node
are valid SAW-trees, and contain bounding box information for the
first and second halves of the SAW, etc., until the leaves of the tree
store individual sites. 
The SAW-tree is related to the R-tree~\cite{Guttman1984CS}, a data
structure used in the field of computational geometry, but with additional
information encoding the state of the SAW. 
Thus far the SAW-tree has been implemented for $\Z^d$, but
can be straightforwardly adapted to other lattices and the continuum,
as well as other polymer models with short-range interactions.
To guarantee optimal performance, we implement the SAW-tree so that it
is balanced, i.e. so that the  
depth never exceeds some fixed constant times $\log N$. We define the
\emph{level} of a node as the number of
generations between a node and the leaves.

\begin{figure}[!ht]
\resizebox{3.375in}{!}{\includegraphics{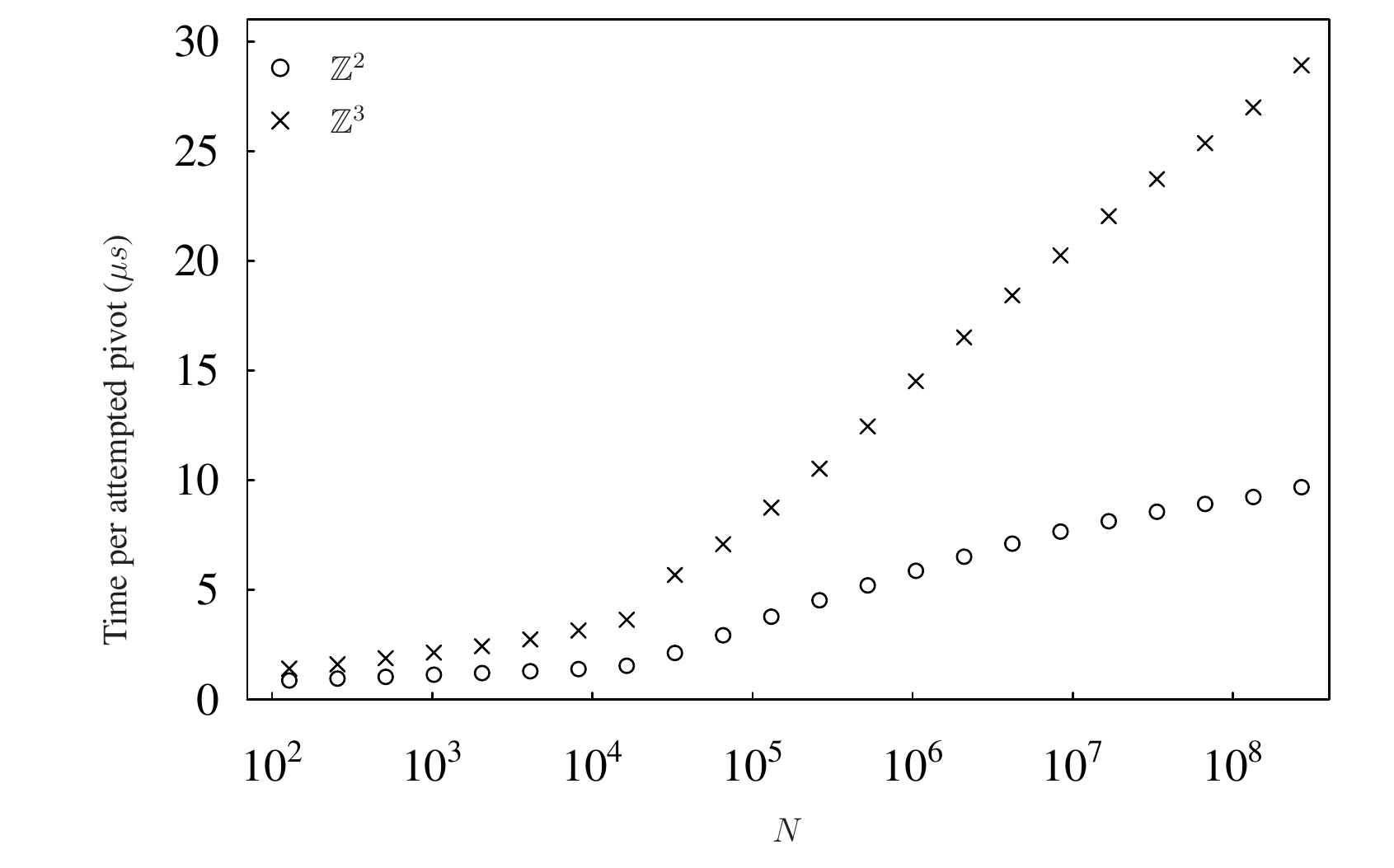}}
\caption{$T(N)$ for $\Z^2$ and
  $\Z^3$. Note that these estimates were obtained in a separate data run
  on a different computer from the main experiment, with lengths from
  $N = 2^7 - 1$ to $N = 2^{28} - 1$.} 
\label{fig:cputimes}
\end{figure}

Bounding boxes enable us to rapidly determine if two sub-walks
intersect after a pivot attempt: if two
bounding boxes do not intersect, then the sub-walks which they contain
cannot intersect.
If a pivot attempt is successful, then it is necessary to
resolve all intersections between bounding boxes of different nodes in
the tree on opposite sides of the pivot site. 
Our implementation
ensures that intersection tests are typically performed between
bounding boxes of nodes which are at the same level.
We argue in \cite{Clisby2009bSAW}
that the nodes at fixed level in the SAW-tree form a renormalized
walk, and the intersections between bounding boxes correspond to
contacts in the original walk. This implies that at each level there
are $O(1)$ intersections, and as the tree has $O(\log N)$ levels this
leads to the conclusion that a successful pivot takes time $O(\log N)$.
Successful pivots occur with probability
$O(N^{-p})$, so overall mean time spent on successful pivots is
$O(N^{-p} \log N)$. 
When a pivot attempt is unsuccessful, with high probability the first
intersection occurs near the pivot site. Thus only a small
fraction of the SAW-tree needs to be traversed to find the
intersection, and we argue in \cite{Clisby2009bSAW} that this takes
mean time $O(1)$.  
Unsuccessful pivots occur with probability
$O(1)$, and so the overall behavior is  $T(N) = O(N^{-p}
\log N + 1) = O(1)$. 
In Fig.~\ref{fig:cputimes} we show $T(N)$ for
$\Z^2$ and $\Z^3$ from a separate data run, with maximum length $N =
2^{28} - 1 \approx 2.68\e{8}$. 
In both cases it is apparent 
there is a crossover 
due to the shorter latency of cache versus main memory.
In \cite{Clisby2009bSAW} we argue that O(1) behavior may be
reached only for very large $N$, which makes interpretation of
Fig.~\ref{fig:cputimes} difficult.
For $\Z^2$ some curvature is visible,
and the trend appears consistent with $T(N)$ approaching a constant
for sufficiently large $N$. The exponent $p$ is smaller for $\Z^3$ ($p
\approx 0.11$) compared with $\Z^2$ ($p \approx 0.19$); hence,
the approach to a constant is far slower, and in fact
almost no curvature is visible for $\Z^3$.
We believe the numerical evidence provides a strong case that $T(N)$
is at most $O(\log N)$, and is consistent with $T(N) = O(1)$; see
\cite{Clisby2009bSAW} for more details.

$T(N)$ is shown for the various implementations
in 
Table \ref{tab:performance}. For SAWs of length $N=10^6$ on the cubic
lattice, the performance gain for our implementation is
approximately 200 when compared with Kennedy's, and over
a thousand
when compared with that of
Madras and Sokal~\footnote{These numbers are intended as
  a rough guide, as they are machine and compiler dependent.}. The
dramatic performance gain from the new implementation
not only makes
it possible to obtain large samples of walks with millions of steps, it also makes
the regime of very long walks, of up to $10^9$ steps, accessible to 
computer experiments. 
\begin{table}[!b]
\caption{$T(N)$, mean time per attempted pivot for $N$-step SAWs.}
\label{tab:performance}
\begin{ruledtabular}
\begin{tabular}{cccc}
Lattice  & Madras and Sokal & Kennedy & This Letter \\ \hline
Square & $O(N^{0.81})$ & $O(N^{0.38})$ & $O(1)$\\
Cubic & $O(N^{0.89})$ & $O(N^{0.74})$ & $O(1)$\\
\end{tabular}
\end{ruledtabular}
\end{table}

For SAWs of length $N$, it is expected that the exponential
autocorrelation time is approximately $O(N/f)$~\cite{Madras1988SAW}, where
$f$ is the fraction of
pivot attempts which are successful.
The first $20 N/f$ configurations were discarded, ensuring
that for all practical purposes SAWs were sampled from the uniform
distribution. Batch 
estimates of $\langle  
R_{\mathrm{e}}^2 \rangle_{33554431}$,
using a batch size of $10^8$, are shown in
Sec.~4 of \cite{Clisby2009bSAW}; in this case
the first 50 
batches were discarded, 
while initialization bias is visually apparent for (at most) the
first 10 batches.

The computer experiment was performed on a cluster of AMD
Opteron Barcelona 2.3GHz quad core processors,
for a total of 16500
CPU hours. Code was written in C, and compiled with gcc.
$10^{11}$ pivot attempts were made on SAWs of length ranging from 15 to
$3.36\e{7}$, for a grand total of $1.89
\times 10^{13}$ pivot attempts.
Data were collected from every pivot attempt for $f$ and the
Euclidean-invariant moments $R_x^{2k} = \left(R_x^2\right)^k$ with $x \in \{\mathrm{e,g,m}\},
1 \leq k \leq 5$ 
\footnote{Data from 
  $R_x^{2k}$ with $2 \leq k \leq 5$ have
  not been analyzed here.
  These data are inferior for critical exponent
  estimates, but may in the future be used to calculate universal
  amplitude ratios.}.
The longest walks with
$N=3.36\e{7}$ required 3GB of memory; much longer
walks could conceivably be simulated in the future.
By comparing fits from the whole data set ($N \leq 3.36\e{7}$) with fits
from a reduced
data set ($N \leq 2.1\e{6}$), we confirmed that
data from the longest walks were indeed highly useful in tying down
the various estimates (see Fig.~1 in
\cite{Clisby2009aSAW}). However, the greatest
benefit from the simulation of truly
long walks, of say $10^9$ steps, may be the ability to directly
simulate properties of realistic
systems, such as DNA knotting, rather than determination of
universal parameters. 

Monte Carlo estimates of global parameters are collected in Tables
II-V, Sec. 2 of \cite{Clisby2009aSAW}, with confidence intervals
calculated using the standard
binning technique.
For all lengths
studied the integrated autocorrelation time of the Markov chain is much less
than the batch size of $10^8$. 
We confirm the accuracy of the confidence interval estimates by studying
the effect of batch size in Sec.~5 of \cite{Clisby2009bSAW}.

\section{Analysis}

We estimated the critical exponents $\nu$ and $\Delta_1 \approx 1/2$ and
associated amplitudes $D_x$ and $b_x$ by fitting the leading
term  and 
leading correction of 
Eq.~(\ref{eq:asymptotic}) via weighted non-linear
regression.
We truncated the data set by requiring $N \geq N_{\mathrm{min}}$,
with $N_{\mathrm{min}}$ a free parameter. We shifted the value of $N$
of Eq.~\ref{eq:asymptotic} by an 
amount $\delta N_x$ to obtain smoother
convergence by altering the sub-leading corrections (see
e.g. \cite{Clisby2007SAW}); estimates for $\nu$, $\Delta_1$, $D_x$, and
$b_x$ are unaffected in
the limit $N_{\mathrm{min}} \rightarrow \infty$. 
With $\delta N_{\mathrm{e}}= .35$, $\delta N_{\mathrm{g}}=1$,  
$\delta N_{\mathrm{m}}=0.4$,
our final model was 
\begin{align}
\langle R_x^2 \rangle_N &= D_x (N+\delta N_x)^{2\nu} \left[1 +
  \frac{b_x}{(N+\delta N_x)^{\Delta_1}} \right],
\label{eq:model}
\end{align}
Unfortunately we cannot fit the next-to-leading corrections
with exponents $1$, $\Delta_2 \approx 1$,  
and $2\Delta_1 \approx 1$ as the differences
between them are far too small to resolve.
For sufficiently large
$N_{\mathrm{min}}$ we found that reduced $\chi^2$ values for all fits
approached 1 from above, indicating Eq.~\ref{eq:model} is asymptotically
correct.

\begin{figure}[!b]
\resizebox{3.375in}{!}{\includegraphics{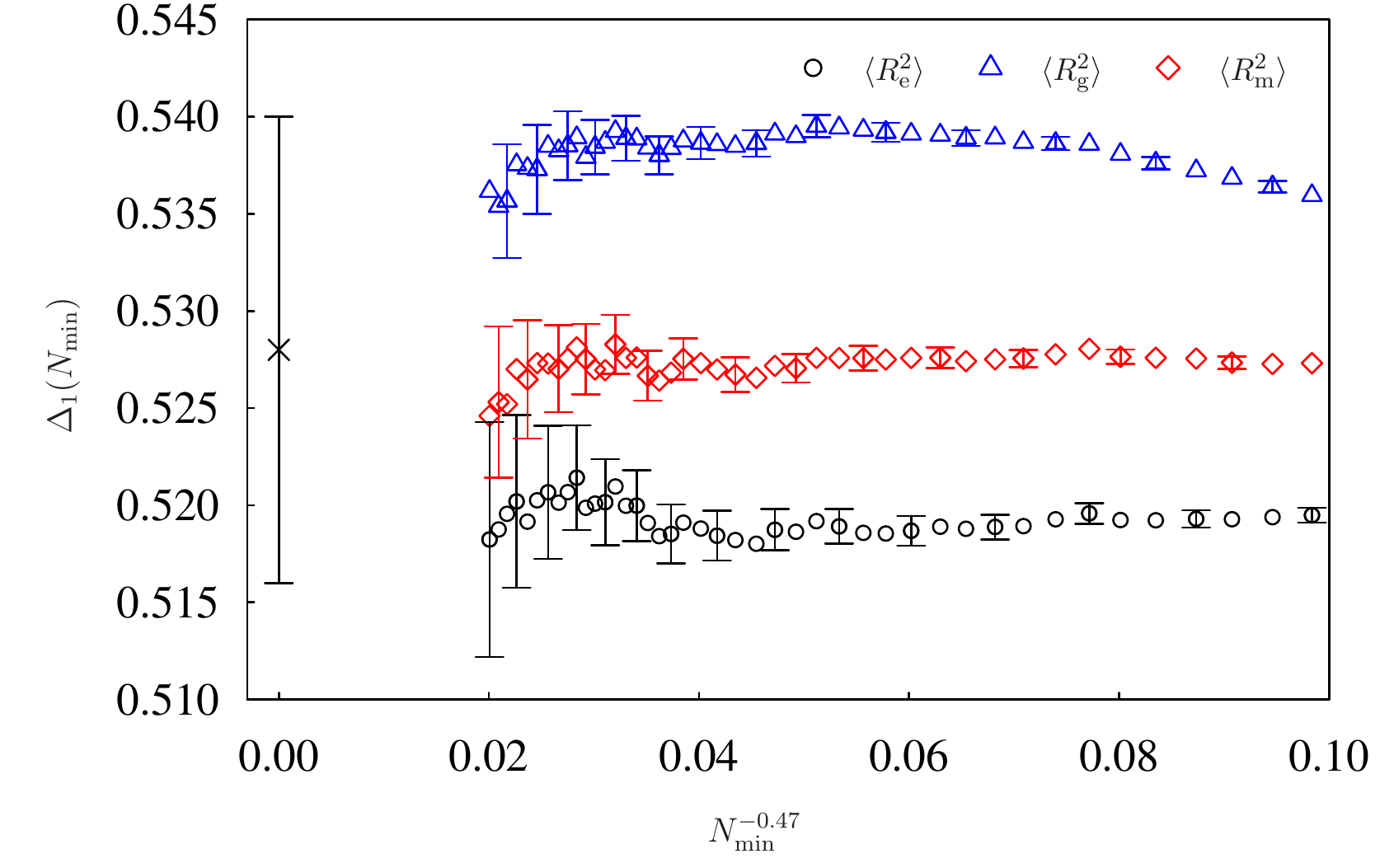}}
\caption{Estimates for $\Delta_1$.}
\label{fig:deltafit}
\end{figure}

\begin{figure}[!ht]
\resizebox{3.375in}{!}{\includegraphics{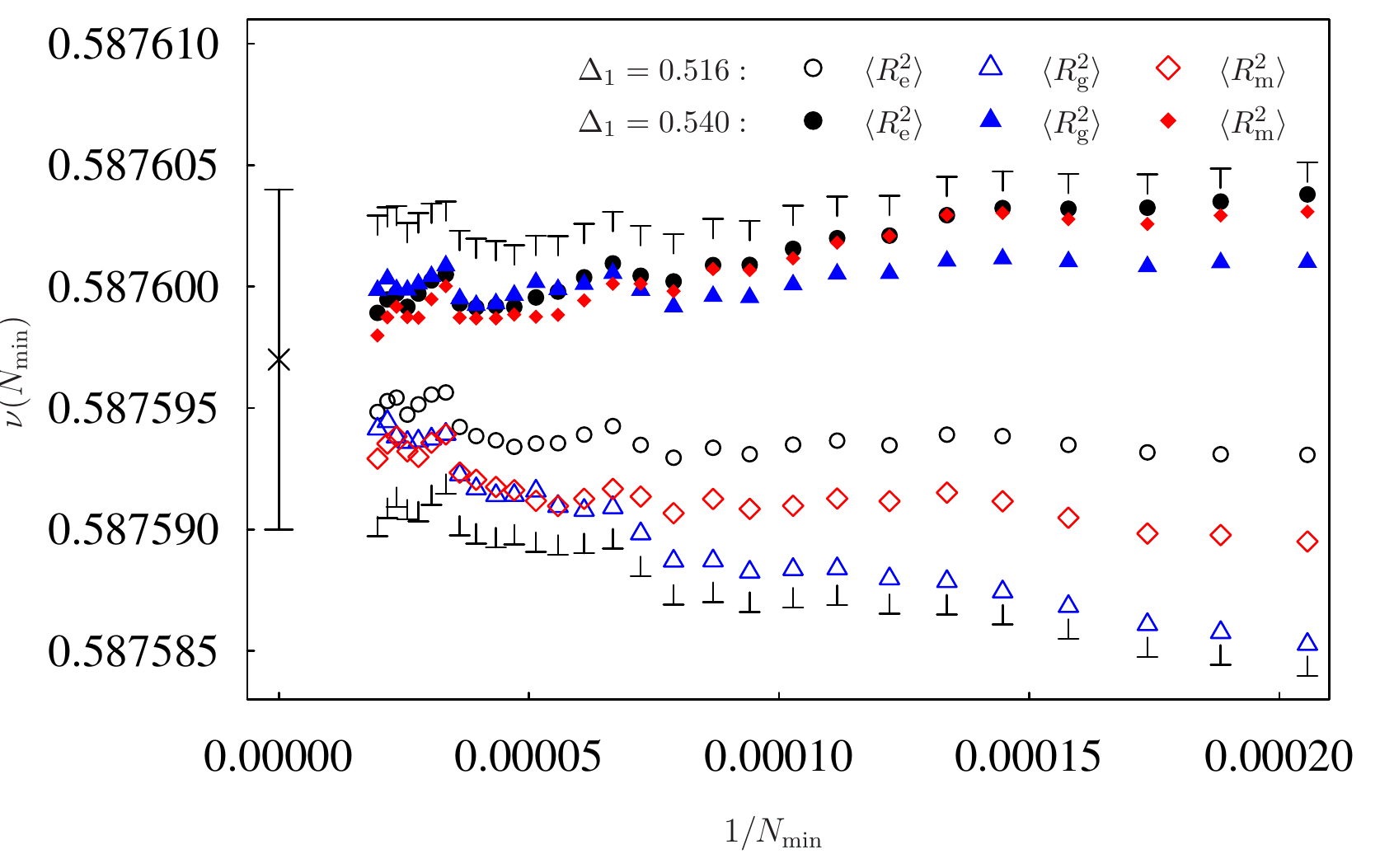}}
\caption{Estimates for $\nu$ from fits with biased $\Delta_1$. 
We show
  the envelope of
  maximum and minimum values of endpoints of error bars for all observables.}
\label{fig:nufitbias}
\end{figure}

Final estimates of parameters have been made directly from
Figs. \ref{fig:deltafit} and \ref{fig:nufitbias}, combining multiple
sources visually in an
attempt to make estimates robust, and allow the reader to critically
evaluate our final results. We do not distinguish
between \emph{subjective} and
\emph{statistical} errors, as we believe that in this context the
distinction is itself quite
subjective~\footnote{e.g., what value of $N_{\mathrm{min}}$
  should be chosen for the statistical error?}.
We provide here some guidance for the interpretation of Figs.
\ref{fig:deltafit} and \ref{fig:nufitbias}, and refer the interested
reader to \cite{Guttmann1989Books} for (much) more information on series
analysis.
\begin{itemize}
\setlength{\itemsep}{1pt}
\setlength{\parskip}{0pt}
\setlength{\parsep}{0pt}
\item{We plot estimates against $N_{\mathrm{min}}^{-y}$, where $y$ is chosen such that
  $N_{\mathrm{min}}^{-y}$ is
  of the same order as the residual error from the fit. The estimates for
  $\Delta_1$ and $b_x$ have
  $y = 1 - \Delta_1 \approx 0.47$, and $y=1$ for $\nu$ and $D_x$.}
\item{We seek to extrapolate the fits to $N_{\mathrm{min}}=\infty$, or $N_{\mathrm{min}}^{-y} =
  0$. Depending upon whether the true value $y_{\mathrm{exact}}$ is less than,
  equal to, or greater than $y$, the estimates would approach a
  limiting value at $N_{\mathrm{min}}^{-y} = 0$ with infinite, finite, or zero slope
  respectively. }
\item{Successive estimates are highly correlated, and so any
  trend which lies within the error bars should be disregarded. Only
  some of the error bars are plotted in order to reduce
  visual noise.}
\item{There are no bounds on the errors of the truncated asymptotic
  formulae, and hence the interpretation of the graphs is subjective.
  The underlying systematic error is observable dependent, and so
  combining estimates from a variety of observables improves
  robustness.}
\end{itemize}

In Fig.~\ref{fig:deltafit} we plot estimates of $\Delta_1$ with our
final result plotted at 0; the error bar reflects the scatter
between observables.
In Fig.~\ref{fig:nufitbias} we plot
estimates for $\nu$, biased with the lower and upper limits of our
range for $\Delta_1$, with our final result at 0. Similar plots for
the amplitudes are given in Figs. 5 and 6 of \cite{Clisby2009aSAW}.
We have conservatively chosen the error bar for the final result to
encompass estimates from all observables
$\langle R_x^2 \rangle$.
As the
amplitudes $D_x$ are highly correlated
with estimates for $\Delta_1$, the biasing of $\Delta_1$ greatly
extends the range over which stable fits can be obtained. This is the
reason the biased fits are preferred over the unbiased fits shown in
Fig.~3 of \cite{Clisby2009aSAW}.

\section{Results}

We report our final results in Table~\ref{tab:parameters}, and in
addition we have $D_{\mathrm{m}} = 0.58687(12)$, $b_{\mathrm{e}} =
-0.49(5)$, $b_{\mathrm{g}} = -0.1125(125)$, and $b_{\mathrm{m}} = -0.295(30)$.
If one assumes the hyperscaling relation $d \nu = 2 - \alpha$, then
one also obtains $\alpha = 0.237209(21)$.
The estimates of
$\nu$, $D_{\mathrm{e}}$, and $D_{\mathrm{g}}$ are in accordance with
previous results, 
although considerably more accurate.
The estimate for $\Delta_1$ is
more accurate than the previous Monte Carlo
value \cite{Li1995SAW}, but less accurate than the Monte Carlo
renormalization group
estimate of \cite{Belohorec1997SAW}, which relies upon an uncontrolled
although accurate approximation. The claimed accuracy of the
field theory estimates~\cite{Guida1998SAW} for $\Delta_1$ is also
comparable, but as
discussed by Li et al.~\cite{Li1995SAW} these calculations have
underlying systematic errors of uncertain magnitude.
Any desired amplitude ratios can be calculated from the amplitude
estimates. The rational number with smallest denominator within 3
standard deviations of
$\nu=0.587597$ is $161/274$,
suggesting that $\nu$ cannot be
expressed as a rational number with small denominator.

\begin{table}[!ht]
\caption{Comparison of parameter estimates.}
\label{tab:parameters}
\begin{ruledtabular}
\begin{tabular}{lllll}
\multicolumn{1}{c}{Source\footnote{Abbreviations: MC $\equiv$ Monte
    Carlo, FT $\equiv$ Field theory, $d=3$ $\equiv$ $d=3$ expansion,
    $\epsilon$ bc $\equiv$ $\epsilon$ expansion with boundary
    conditions, MCRG $\equiv$ Monte Carlo renormalization group.}} & \multicolumn{1}{c}{$\nu$} & \multicolumn{1}{c}{$\Delta_1$} &
  \multicolumn{1}{c}{$D_{\mathrm{e}}$} & \multicolumn{1}{c}{$D_{\mathrm{g}}$}
   \\ \hline
This Letter & 0.587597(7) & 0.528(12)& 1.22035(25)& 0.19514(4)\\
\cite{Clisby2007SAW}\footnote{Using Eqs. (74) and (75) with $0.516 \leq \Delta_1
  \leq 0.54$.} Series  &   0.58774(22) &
&1.2178(54) &\\
\cite{Prellberg2001SAW} MC & 0.5874(2)  &   & & \\
\cite{MacDonald2000SAW}\footnote{No error
  estimates were made in \cite{MacDonald2000SAW}, but estimates for
  $\nu$ were in the 
range $0.5870 \leq \nu \leq 0.5881$.} Series & 0.58755(55)  & & 1.225 &\\
\cite{Guida1998SAW} FT $d=3$& 0.5882(11)  & 0.478(10)  &  &  \\
\cite{Guida1998SAW} FT $\epsilon$ bc & 0.5878(11)& 0.486(16)  & &   \\
\cite{Belohorec1997SAW} MCRG & 0.58756(5)   & 0.5310(33)   & & \\
\cite{Li1995SAW}\footnote{In addition $b_{\mathrm{e}}=-0.483(39)$, $b_{\mathrm{g}} =
  -0.1143(47)$. $D_{\mathrm{e}}$, $D_{\mathrm{g}}$, $b_{\mathrm{e}}$, and $b_{\mathrm{g}}$ estimates were biased
  with $\nu=0.5877$,
  $\Delta_1=0.56$; the confidence intervals were not
  intended to be taken seriously.} MC   &  0.5877(6) & 0.56(3) & 1.21667(50) & 0.19455(7) \\
\end{tabular}
\end{ruledtabular}
\end{table}

We would like to stress that, due to the neglect of sub-leading terms,
there are underlying
systematic errors in our estimates
which \emph{are not} and
\emph{cannot be} controlled. We have the luxury of high quality data
from long walks, and have attempted to be conservative with
our claimed errors, but acknowledge there is a risk that the 
(subjective) confidence intervals may not be sufficiently large.
 
\section{Conclusion}

In summary, an efficient version of the pivot
algorithm for SAWs has been implemented and used to calculate $\nu$;
the algorithms 
developed promise to be widely useful in the Monte Carlo
simulation of SAWs and related models of polymers.

\begin{acknowledgments}
I thank I.~G.~Enting, A.~J.~Guttmann, G.~Slade, A.~Sokal, and an
anonymous referee for useful
comments on the manuscript.
Computations were performed using the
resources of VPAC.
Financial support from the Australian Research Council is gratefully
acknowledged. 
\end{acknowledgments}

\FloatBarrier

\end{document}